\documentclass{article}

\usepackage{amsmath}
\usepackage{booktabs} 
\usepackage[ruled,vlined]{algorithm2e}
\usepackage{xcolor}
\usepackage{graphicx}
\usepackage{tikz}
\usetikzlibrary{bayesnet}
\usepackage{siunitx}
\usepackage[parfill]{parskip}
\usepackage{authblk}
\newtheorem{assumption}{Assumption}

\newtheorem{lemma}{Lemma}

\newcommand{\ra}[1]{\renewcommand{\arraystretch}{#1}}

\author[1]{Ercan Yildiz, Joshua Safyan, and Marc Harper\thanks{Authors can be reached at: meyildiz@gmail.com, jsafyan@gmail.com, and marc.harper@gmail.com. We are grateful to Micha Segeritz and David Sinclair for their feedback on this work.} \\ Google, Mountain View, CA}

\title{User Sentiment as a Success Metric: \\ Persistent Biases Under Full Randomization}

\begin{document}
\maketitle

\begin{abstract}
We study user sentiment (reported via optional surveys) as a metric for fully randomized A/B tests. Both user-level covariates and treatment assignment can impact response propensity. We propose a set of consistent estimators for the average and local treatment effects on treated and respondent users. We show that our problem can be mapped onto the intersection of the missing data problem and observational causal inference, and we identify conditions under which consistent estimators exist. We evaluate the performance of estimators via simulation studies and find that more complicated models do not necessarily provide superior performance.
\end{abstract}

\section{Introduction}
Experimentation via A/B testing is widespread practice especially among technology companies~\cite{Xu:2015:ICA:2783258.2788602,Kohavi:2013:OCE:2487575.2488217,Tang:2010:OEI:1835804.1835810}. A typical A/B test splits a population in two or more distinct groups at random, exposing each to different experiences. The experiment arm is then evaluated by a set of predefined metrics~\cite{Tang:2010:OEI:1835804.1835810}. Engagement-based metrics are frequently used, and the entire process can be automated either by in-house infrastructure~\cite{Tang:2010:OEI:1835804.1835810,Xu:2015:ICA:2783258.2788602} or by third party services~\cite{optimizely}. 

Organizations have also measured user sentiment and attitudes at scale since the late 1980s~\cite{10.2307/184036, nps, CSAT-retention}, mainly driven by the relationship between consumer attitudes, long term loyalty, and revenue. Sentiment changes in an undesirable direction have also been shown to result in decreased engagement and lower long-term retention~\cite{CSAT-retention}.  

\begin{figure}[h!]\label{fig:hats}
 \centering
\includegraphics[width=0.7\textwidth]{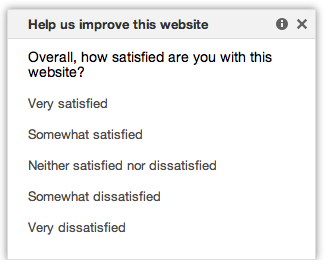}
\caption{Sample UI based survey}
\end{figure}

Improvements in survey technology have enabled faster and more targeted feedback. Historically, phone and mail surveys as well as in-person studies were the primary source of consumer data. Email-based surveys brought increased reach and decreased the cost of collecting user feedback. More recently, companies have brought surveys into UIs and started capturing user sentiment during the experience~\cite{Muller:2014:HLI:2686612.2686656}. An example of a UI-based survey is given in Figure~\ref{fig:hats}.

In recent years, companies have begun merging experimentation and user sentiment by utilizing sentiment as a success metric for product launches. This has become possible due to three major advantages of UI-based surveys over traditional survey mediums 1) mapping survey responses to treatment exposure is straightforward as surveys are attached to experiences, 2) capturing user sentiment in close temporal-proximity to the actual user experience increases validity of responses, 3) UI surveys have the potential to reach many users and have higher response rates compared to traditional channels~\cite{Muller:2014:HLI:2686612.2686656}.

Employing sentiment-based metrics in an A/B setting has its own challenges. First, unlike a typical engagement metric, sentiment data is incomplete in most cases. In particular, we have the ability to observe levels of engagement across all users in an A/B test\footnote{Clearly, some users may not engage, though this is an observation of no engagement and differs from survey non-response.} while sentiment is measured through optional surveys to which not all users respond. Second, response behavior may be correlated with sentiment as well as treatment exposure. As a result, naive comparisons of sentiment-based metrics risk biased conclusions.

Accordingly, we focus on a scenario where user sentiment reported via optional surveys is used as a success metric for a fully randomized A/B test. There exists a set of user-level confounders which might impact the sentiment as well as the propensity to respond to the survey. Moreover, we allow treatment/control assignment to impact the probability of response. Our goal is to estimate both the population-level average treatment effect (ATE), and the local treatment effect on users that were treated and responded to the survey (ATETR).

\section{Related Work and Our Contribution}
Our setting relates to the well-known missing data and survey non-response problems \cite{brick1996handling, kalton2003weighting, kalton1983compensating}. Missing data literature focuses on the problem of estimating a population statistic from an incomplete dataset~\cite{Little:1986:SAM:21412, 10.2307/2335739, kang2007, doi:10.1111/j.1541-0420.2005.00377.x}. In this scenario, there is no experimentation; only a single population with missing observations~\cite{kang2007}. Typically, the data is assumed to be missing at random (MAR), i.e., missing data can be fully explained by a set of observed variables~\cite{10.2307/2335739}. Under the MAR assumption, a rich set of consistent estimators have been proposed including outcome regression, inverse propensity weighing, and various doubly-robust methods~\cite{kang2007, doi:10.1111/j.1541-0420.2005.00377.x}. Though related, our problem departs from the traditional missing data problem because we have two distinct groups that are exposed to different experiences which impact both outcomes and the likelihood of missing data.

Our problem also intersects with causal inference, where one is interested in estimating the effect of a treatment using observational and/or quasi-experiment data~\cite{doi:10.1111/j.1541-0420.2005.00377.x, 2018arXiv181007406O, hainmueller_2012, doi:10.1111/ectj.12097, imbens_rubin_2015, Pearl:2009:CMR:1642718}. In the absence of a randomized controlled trial (RCT), the relationship between the treatment assignment and the outcome of interest may be confounded. The causal inference literature includes a wide range of methods to adjust for confounding in treated and non-treated populations based on a set of observed variables~\cite{doi:10.1111/j.1541-0420.2005.00377.x, 2018arXiv181007406O, doi:10.1111/ectj.12097, Pearl:2009:CMR:1642718}. While we rely heavily on causal inference, our setting also deviates from a typical causal problem. In particular, treatment assignment in our case is fully randomized, so there is no treatment selection bias. However, since responses are optional, there is still non-response bias. 

We propose a framework at the intersection of missing data and causal inference to adjust for differential response bias in surveys in the A/B setting. We map ATE estimation into a missing data problem across two (or more) populations, and ATETR estimation into a causal inference problem by leveraging methods for estimating the average treatment effect on the treated (ATET). Additionally, we provide a simulation study using existing models to demonstrate a stylized application of our method. Finally, in the case of ATETR, we argue that a simple regression based covariate control may provide comparable performance to more complicated and resource heavy techniques.

\section{Problem setup}
Consider a scenario where user sentiment is used to evaluate a treatment. For the sake of simplicity, we assume that sentiment is collected via a single question{\footnote{It is straightforward to expand our results to the case where sentiment is measured by a series of questions.}}. For instance, if we are interested in user satisfaction we may ask \textit{How satisfied are you with the overall experience?}. A user might pick one of the following $\{$ {\textit{Extremely Satisfied, Satisfied, Neither Satisfied Nor Dissatisfied, Dissatisfied, Extremely Dissatisfied}} $\}$\cite{doi:10.1177/109467059800100102, anderson_1997}. It is common practice to code favorable responses higher, e.g, "Extremely Satisfied" is commonly coded as $5$ and "Extremely Dissatisfied" is coded as $1$. Additionally, this scale can be simplified using a Top 2 Box score, i.e, $\{4,5\}$ (\textit{Satisfied, Extremely Satisfied}) are mapped to $1$ and $\{1,2,3\}$ are mapped to $0$ \cite{10.2307/30162061, myers1999measuring}. We focus on Top 2 Box binary sentiment measures, though our proposed approach generalizes across different sentiment measures and coding practices. Furthermore, we restrict to the case where survey response is optional, which differentiates our problem from a typical RCT\footnote{If we could mandate survey response we would simply compare the averages of the treatment and control groups to calculate the average treatment effect.}.

Assume that there are $N$ individuals in a population, and the population is randomly split into treatment and control groups. We will assume that all of the individuals in the population are exposed to either a treatment or a control condition. 
$T_{i}$ is a dummy variable indicating whether the individual $i$ is exposed to treatment or control, i.e, $T_i=1$, if individual $i$ is exposed to treatment, and $T_i=0$ if individual is exposed to control. $(Y_i(1),Y_i(0))$ captures the (scalar) sentiment measure for individual $i$ under treatment and control conditions, respectively. Finally, $(\Delta_i(1),\Delta_i(0))$ are indicators for survey response behavior under treatment and control conditions. In other words, for a user assigned to condition $T_{i}$, $\Delta_i(T_i)$ will be equal to $1$ if the user responds to the survey. 

We are interested in estimating two distinct quantities: the average treatment effect on the overall population (ATE), i.e:\begin{align}\label{main_eq}
    E[Y(1)-Y(0)],
\end{align}
and the average treatment effect on individuals who are exposed to the treatment and respond to the survey (Average Treatment Effect on Treated and Responded - ATETR), i.e:
\begin{align}\label{main2_eq}
    E[Y(1)-Y(0)| T=1, \Delta(T) = 1].
\end{align}
The former resembles an engagement metric, which is observed across all individuals, by aiming to provide population level impact. Given the wide adoption of engagement-based metrics for product evaluations, comparability and consistency with such metrics is an important factor. However, stricter assumptions are required to extrapolate from the observed responses to the overall population. Such assumptions may be problematic given that average online surveys have a response rate lower than 10$\%$~\cite{10.2307/24311411}. While the latter measure does not require estimating the sentiment of survey non-respondents, it is more difficult to compare to engagement type metrics since the ATETR estimates a local effect rather than a population effect.

Going forward, we utilize two assumptions depending on the measure of interest.

\begin{figure}[]
  \centering
\includegraphics[width=0.7\textwidth]{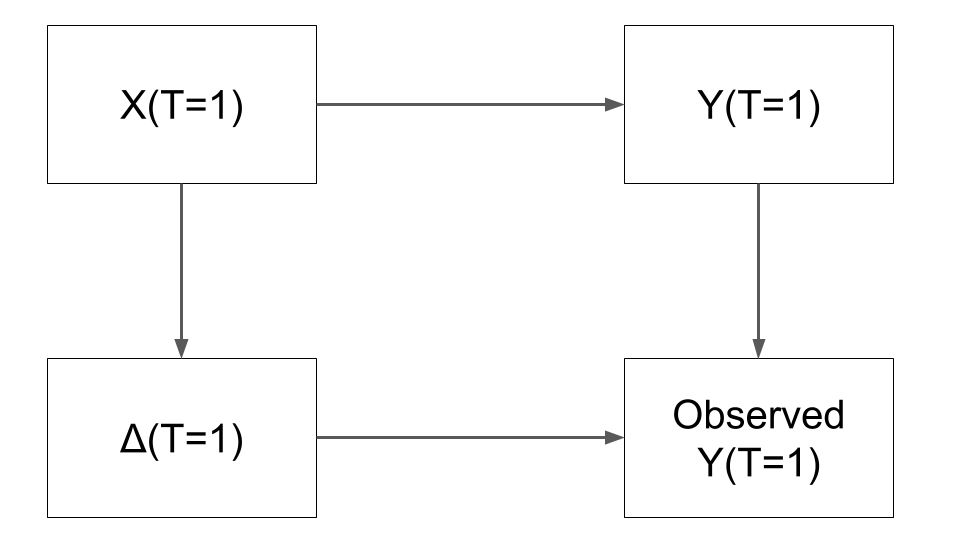}
\caption{Causal graph representing Assumption~\ref{assumption_1}. $X(T)$ influences both user sentiment $Y(T)$ and propensity to respond $\Delta(T)$. There is no other path between user sentiment and propensity to respond. Therefore, $Y(T)$ and $\Delta(T)$ are independent conditioned on $X(T)$.}
\label{figure_causal_1}
\end{figure}

\begin{assumption}[Missing At Random]\label{assumption_1} 
There exists a set of observed variables $X_i(T_i)$, $T_i \in \{0,1\}$ such that:
\begin{align}
\Delta_i(T_i) \perp Y_i(T_i) ~|~ X_i(T_i), \label{assumption1_eq1} \\
P(\Delta_i(T_i) = 1 ~| ~ X_i(T_i)) > 0  \label{assumption1_eq2}
\end{align}
$\forall ~ i \in \mathcal{N} = \{1,\ldots, N\}$. 
\end{assumption}
The first equation of Assumption \ref{assumption_1} is known as Missing At Random (MAR), and says that systematic differences between the missing and observed values can be entirely explained by a set of observed variables~\cite{10.2307/2335739, Little:1986:SAM:21412, kang2007}. The second equation of the assumption requires that the probability of observing complete data is always positive for a given individual $i$. Since $X_i(T_i)$ is a function of the assignment, i.e, $T_i$, it is possible that different sets of variables can capture confoundedness in treatment and control groups. We present the causal graph representing Assumption~\ref{assumption_1} in Figure~\ref{figure_causal_1}. The equivalent of the causal graph for $T=0$ can be constructed simply by replacing $(1)$'s with $(0)$'s. 

\begin{assumption}
\label{assumption_2} There exists a set of observed variables $X_i$ such that:
\begin{align}
(\Delta_i(0),\Delta_i(1))   \perp Y_i(0) ~|~ X_i \label{assumption2_eq1} \\
P(\Delta_i(0) = 1 ~ | ~ X_i) > 0, P(\Delta_i(1) = 1 ~ | ~ X_i) > 0 \label{assumption2_eq2}
\end{align}
\end{assumption}

The first equation of Assumption \ref{assumption_2} requires the joint survey response behaviors to be independent of the sentiment under the control condition given the confounders. We require this assumption to estimate the ATETR, $E[Y(1)-Y(0)| T=1, \Delta(T) = 1]$. Specifically, we want to estimate how those who were assigned to the treatment group and responded to the survey ($Y_i(0), \Delta_i(0) = 1$) would have responded under control. Note that Assumption \ref{assumption_2} does not impose any constraints on the conditional independence of $Y_i(1)$ since we observe the responses from the users who were treated and responded to the survey, thus we only need to estimate the counterfactual sentiment for treated and respondent users under control.

\begin{figure}[]\label{figure_causal_2}
  \centering
\includegraphics[width=0.7\textwidth]{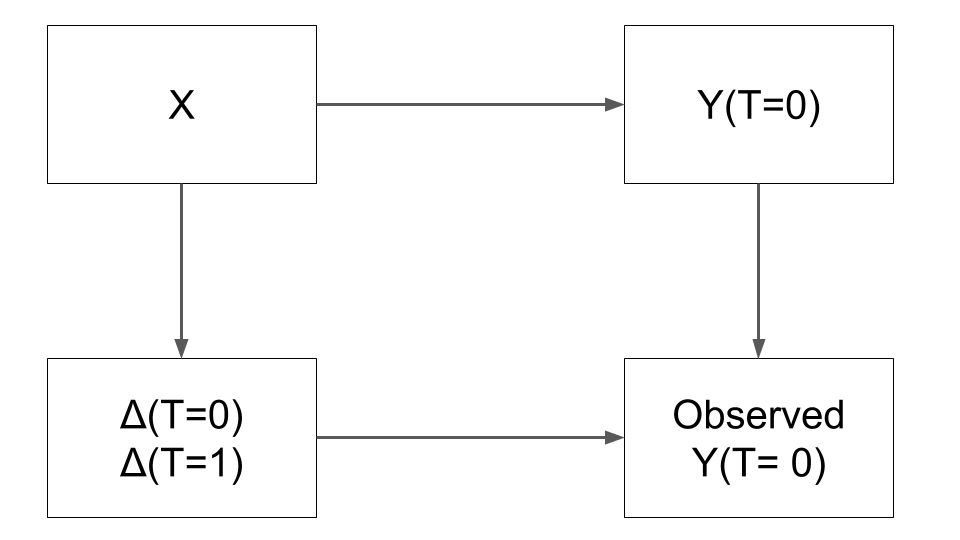}
\caption{Causal graph representing Assumption~\ref{assumption_2}. $X$ influences both user sentiment under control $Y(T=0)$, and propensity to respond under both treatment and control conditions $\Delta(T=0), \Delta(T=1)$. There is no other path between user sentiment and propensity to respond. Therefore, $Y(T=0)$ and $\Delta(T=0),\Delta(T=1),$ are independent conditioned on $X$.}
\end{figure}

We show the causal graph representing Assumption~\ref{assumption_2} in Figure~\ref{figure_causal_2}.

\section{Balancing Treatment and Control}
We propose mechanisms to estimate the measures of interest: ATE and ATETR.
\subsection{Sentiment Average Treatment Effect (ATE)}\label{sub_ate}
We are interested in estimating the Average Treatment Effect on user sentiment over the entire population. Our first results enables parallelization of estimation of ATE by mapping it onto two missing data problems:
\begin{lemma}\label{lemma_ATE}
Let $\theta_1$ denote a consistent estimator of $E[Y(1)|T=1]$, and $\theta_0$ denote a consistent estimator of $E[Y(0)|T=0]$. Then, $\theta_1 - \theta_0$ is a consistent estimator of ATE.
\end{lemma}
The proof of the lemma is straightforward, and follows from the linearity of expectation, linearity of convergence in probability, and randomized treatment assignment \emph{i.e}, $E[Y(1)|T=1]=E[Y(1)|T=0]=E[Y(1)]$. Lemma \ref{lemma_ATE} suggests that one can focus on the treatment and control groups separately and treat each case as an independent missing data problem. Independence results from the fully randomized treatment/control assignment even though survey measures suffer from response bias.  
Lemma \ref{lemma_ATE} is powerful since it enables parallelization of the estimation. 

While we have mapped our ATE estimation into two independent missing data problems, we need further assumptions in order to provide consistent estimators, \emph{i.e.}, $\theta_1,\theta_0$. In particular, we rely on Assumption ~\ref{assumption_1} (also known as  MAR~\cite{doi:10.1111/j.1541-0420.2005.00377.x}) which guarantees consistency for a wide range of estimators including outcome regression (OR), inverse propensity weighting (IPW), and adversarial balancing (AB) estimators\footnote{Additional assumptions for model specifications are required for parametric models.}. 

As discussed in detail in~\cite{doi:10.1111/j.1541-0420.2005.00377.x}, Assumption~\ref{assumption_1} enables us to infer sentiment measures for users who have not responded to the survey in the treatment (control) group by simply utilizing the responses of the users who have responded to the survey in the treatment (control) group.

As an example, we focus on the OR estimator and demonstrate how $\theta_1,\theta_0$ can be calculated. We start with a set of equalities: 
\begin{align}
E[Y(1)] &=  E_{X}[E_{Y}[Y(1)|X]] \nonumber \\
&= E_{X}[E_{Y}[Y(1) |X, T=1, \Delta(1) = 1]] \label{eqATE}
\end{align}
where the first line follows from law of iterated expectation, and the second from Assumption~\ref{assumption_1} and the fact that treatment assignment is random. Equation~\eqref{eqATE} shows that building an estimator $E[Y_i(1) | X_i] = f(X_i)$ trained on observed responses only ($\{ i ~ | T_i =1, \Delta_i(1) =1\}$) provides a consistent estimator for $E[Y(1)]$ provided that $f(.)$ is correctly specified.
\begin{align*}
\theta_1 = |\{i: T(i) =1\}|^{-1}\sum_{i: T(i) =1} f(X_i)
\end{align*}
is the estimator of interest, where $|.|$ is the cardinality of the set. $\theta_0$ can be calculated in a similar fashion. Interested readers can refer to~\cite{doi:10.1111/j.1541-0420.2005.00377.x} for detailed discussions.

We implemented several estimators (including outcome regression) for comparison purposes, and discuss performance in Section~\ref{simulations}.

\subsection{Sentiment Average Treatment Effect on Treated and Responded (ATETR)}
Now we estimate the Average Treatment Effect on Treated and Responded given in Equation~\eqref{main2_eq}. 

Similar to its counterpart in the causal inference literature, \emph{i.e.}, Average Treatment Effect on Treated (ATET), ATETR requires reconstructing counterfactuals for a subset of users. In our case, the subset is those who were assigned to the treatment group and chose to respond to the survey. 
 
There is a consistent estimator for ATETR, defined as follows:
\begin{lemma}\label{lemma_ATETR}
Given the definition of ATETR in Equation~\eqref{main2_eq} and Assumption~\ref{assumption_2}, 
\begin{align}
ATETR &= E[Y(1)-Y(0)| T=1, \Delta(T) = 1] \nonumber \\
&= E[Y(1)| T=1, \Delta(T) = 1] \label{eqn:ATETR0} \\
&- E[Y(0)| T=1, \Delta(T) =1] \label{eqn:ATETR1}
\end{align}
Moreover, $|\{i:T_i=1, \Delta_i(1) = 1\}|^{-1} \sum_{i:T(i)=1, \Delta(i) = 1} Y_i(1)$ is a consistent estimator of Equation~\eqref{eqn:ATETR0}, and Equation~\eqref{eqn:ATETR1} is equal to:
\begin{align}
    E_{X} [E_{X,\Delta(1) = 1} [E_Y[Y(0)| \Delta(0)=1, X] | \Delta(1) = 1, X ]].\label{eqn:ATETR2}
\end{align}
\end{lemma}
Proof of Lemma~\ref{lemma_ATETR} is given in Appendix~\ref{proof_lemma_2}.

As an example, we will once again focus on the OR estimator and demonstrate how ATETR can be calculated.

Equation~\eqref{eqn:ATETR0} can simply be estimated by taking the average sentiment among users that were assigned to the treatment group and responded to the survey ($\{i | T_i = 1,\Delta_i(1)=1\}$). Equation~\eqref{eqn:ATETR1} (or equivalently Equation~\eqref{eqn:ATETR2}) can be estimated by a 3 step process:
\begin{enumerate}
    \item Focus on survey respondents under the control assignment, \emph{i.e.}, $\{ i ~ | T_i =0 , \Delta_i(0) =1\}$, and train a model for  $E[Y_i(0) | X_i] = g(X_i)$.
    \item Construct counterfactual responses for users that were assigned to the treatment and responded to the survey, $E[Y_i(0)|T_i=1, \Delta_i(1)]$ using $g(X_i)$, $\forall~i \in \{i~|T_i=1, \Delta_i(1) = 1\}$. 
    \item Estimate $E[Y(1)| T=1 , \Delta(1) = 1]$ as ${|{i: T(i) =1, \Delta_i(1) =1}|^{-1}} \sum_{i: T_i =1} g(X_i)$
\end{enumerate}

In other words, training a model on the control group and using the trained model to predict counterfactuals over the treatment group provides a consistent estimator. Finally, the difference between the two provides a consistent estimator for ATETR\footnote{Additional assumptions for model specifications are required.}. 

Similar to ATE, we implemented several estimators (including outcome regression) for comparison purposes, and discuss performance in Section~\ref{simulations}.

\section{Simulations}\label{simulations}
\begin{figure}[]
  \centering
\includegraphics[width=0.7\textwidth]{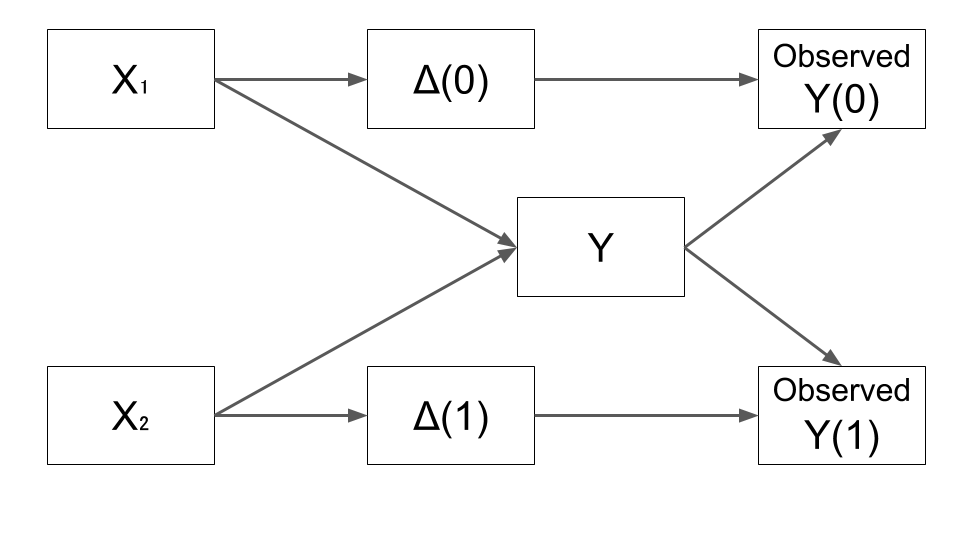}
\caption{Causal graph representing our data generation process. Latent variables ($X = [X_1, X_2]$) generate both individual sentiment $Y$ and response behavior under the treatment and control conditions $\Delta(1), \Delta(0)$.}
\label{figure_simulated_causal}
\end{figure}

We applied our methodology to a simulated dataset. A causal graph representing the generative process is given in Figure~\ref{figure_simulated_causal}. Latent variables ($X = [X_1, X_2]$) generate both individual sentiment $Y$ and response behavior under the treatment and control conditions $\Delta(1), \Delta(0)$. User sentiment $Y$ is the same across treatment/control groups. However, response behaviors $\Delta(0), \Delta(1)$ differ under treatment vs. control and correlate with sentiment, so observed sentiment differs.

Details of the data generation process (DGP) are given in Appendix~\ref{appendix_DGP}. There are $4$ important points regarding the DGP:
\begin{enumerate}
    \item Both Assumptions~\ref{assumption_1} and~\ref{assumption_2} hold. 
    \item There is no difference in the population level sentiment, $E[Y(1)] = E[Y(0)]$.
    \item There are differences in the observed sentiment: 
    \begin{align*}
    E[Y(1) | \Delta (1) = 1] - E[Y(0) | \Delta (0) = 1] \approx  0.07.
    \end{align*}
    \item Response rates under the treatment and control are also different: $E[\Delta(1)] \approx 0.15$, and  $E[\Delta(0)] \approx 0.09$. These numbers correspond to $15\%$ and $9\%$ survey response rates, which are consistent with lower response rates typically observed for online surveys. 
\end{enumerate}

We run $5K$ Monte Carlo simulations with $N=10K$ ($\sim5K$ each in treatment and control). $10K$ is used as the per simulation population size to reflect the small sample sizes typically observed during sentiment studies. Summary statistics are available in Table~\ref{table_sim_param}.

\begin{table*}\centering
\ra{1.2}
\begin{tabular}{lr}
\toprule
\cmidrule(r){1-2}
Metric    & Value \\
\midrule
 $E[\Delta(1)]$ &  $0.15$\\ 
 $E[\Delta(0)]$  & $0.09$\\
 $E[Y(1)-Y(0)]$ & $0$ \\
 $E[Y(1)-Y(0)| T=1, \Delta(1) = 1]$  & $0$ \\
 $E[Y(1)|\Delta(1)=1] - E[Y(0)|\Delta(0)=1]$ & $0.07$\\
\bottomrule
\end{tabular}
\caption{Population level statistics based on GDP in Appendix~\ref{appendix_DGP}.}
\label{table_sim_param}
\end{table*}

\subsection{Sentiment Average Treatment Effect (ATE)}
We implemented four different estimators and compared their performance under two different scenarios.

In scenario 1, we assume that we can observe the actual values of the confounding factors, \emph{i.e.}, $X_1,X_2$. This allows us to correctly specify a model to adjust for all of the confounding variables, and Assumption~\ref{assumption_1} holds.  

Our first estimator is a naive mean comparison among respondents. By definition this is an inconsistent estimator and does not take into account observed confounders. The remaining estimators are based on Lemma~\ref{lemma_ATE}, and use the observed covariates to correct for the underlying bias. We implemented outcome regression (OR)~\cite{doi:10.1111/j.1541-0420.2005.00377.x}, inverse propensity weighing (IPW)~\cite{kang2007}, a doubly robust (DR) estimator based on OR and IPW~\cite{doi:10.1111/j.1541-0420.2005.00377.x}, and an adversarial balancing based estimator (AB) using Logistic Regression as the classifier~\cite{2018arXiv181007406O}.

\begin{figure}[]
  \centering
\includegraphics[width=1\textwidth]{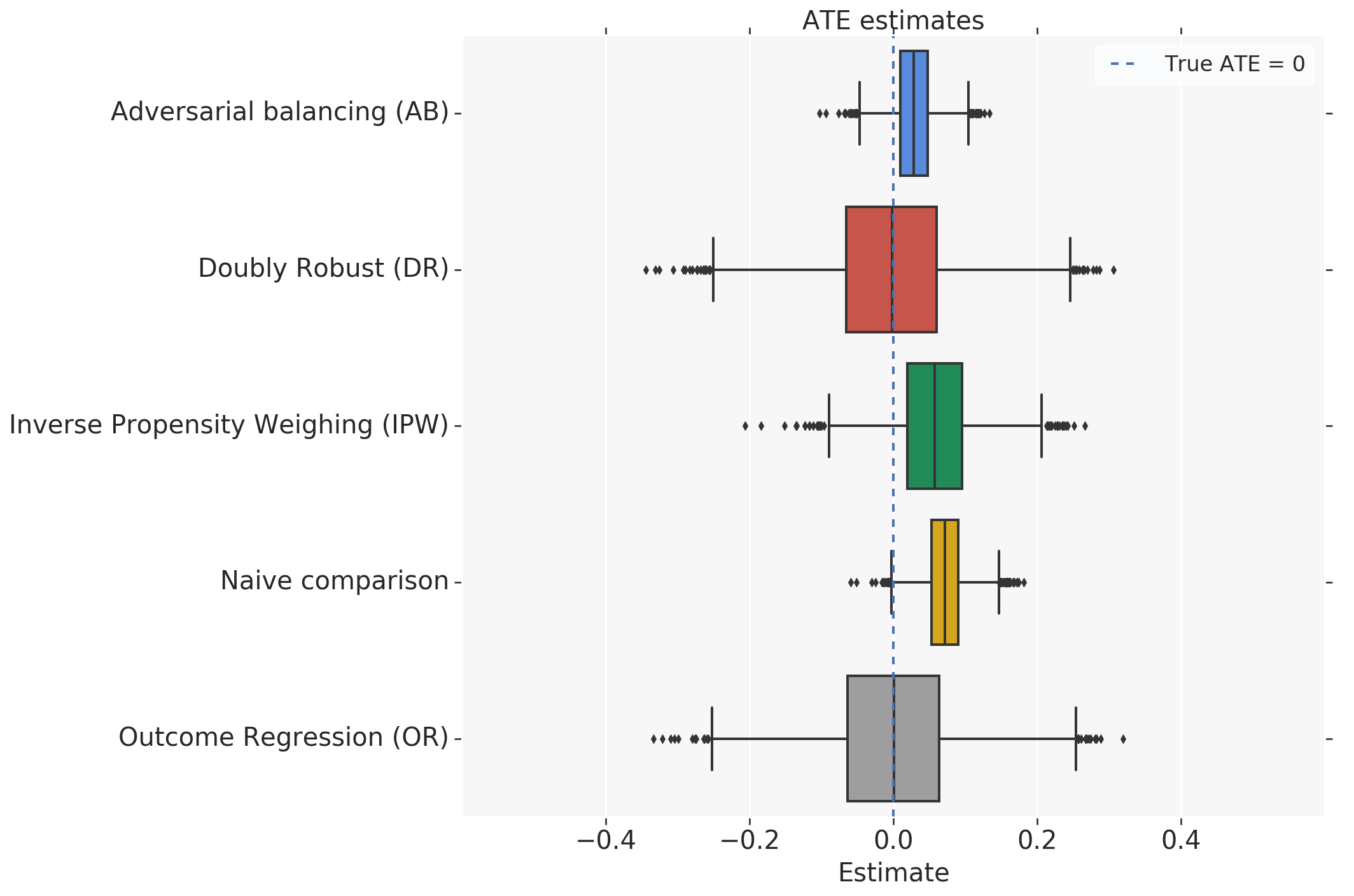}
\caption{Performance of different ATE estimators when true confounders are fully observed. DR and AB has the highest variance.}
\label{figure_sim_ATE_1}
\end{figure}

We measured the performance of each estimator via bias, median absolute error (MAE), and mean squared error (MSE)\footnote{Bias = $E[ATE- \hat{ATE}]$, MAE = $Median(|ATE - \hat{ATE}|)$, and MSE  = $E[(ATE- \hat{ATE})^2]$ } as discussed in~\cite{kang2007}. Results are given in Table~\ref{table_ate_results}. See Figure~\ref{figure_sim_ATE_1} for the distribution of estimates. 

\begin{table*}\centering
\ra{1.3}
\begin{tabular}{llll}
\toprule
\cmidrule(r){1-2}
 Estimator & Bias & MAE & MSE \\
\midrule
 AB & \num{2.8e-2} & \num[math-rm=\mathbf]{2.9e-2} & \num[math-rm=\mathbf]{1.5e-3}\\
 DR & \num{-2.3e-3} & \num{6.1e-2} & \num{8.0e-3}\\
 IPW & \num{5.7e-2} & \num{5.8e-2} & \num{6.2e-3}\\
 Naive comparison & \num{7.1e-2} &  \num{7.1e-2} & \num{5.8e-3} \\
 OR & \num[math-rm=\mathbf]{8.8e-4} & \num{6.1e-2} & \num{8.1e-3} \\
\bottomrule
\end{tabular}
\caption{Performance of different ATE estimators when true confounders are fully observed. Estimators are Adversarial Balancing (AB), Doubly Robust (DR), Inverse Propensity Weighing (IPW), Naive mean comparison and Outcome Regression (OR). The naive estimator has the largest bias while AB estimator has the best MSE performance.}
\label{table_ate_results}
\end{table*}

Naive comparison incorrectly reports a $0.07$ ($7$ percentage points) difference between treatment and control. Variation in the naive comparison is due to the stochasticity of the simulation. The OR estimator has the smallest bias, despite the fact that it suffers from higher variance. IPW has a large bias comparable to the naive estimator. The DR estimator's performance falls in between OR and IPW.

\begin{figure}[]
  \centering
\includegraphics[width=1\textwidth]{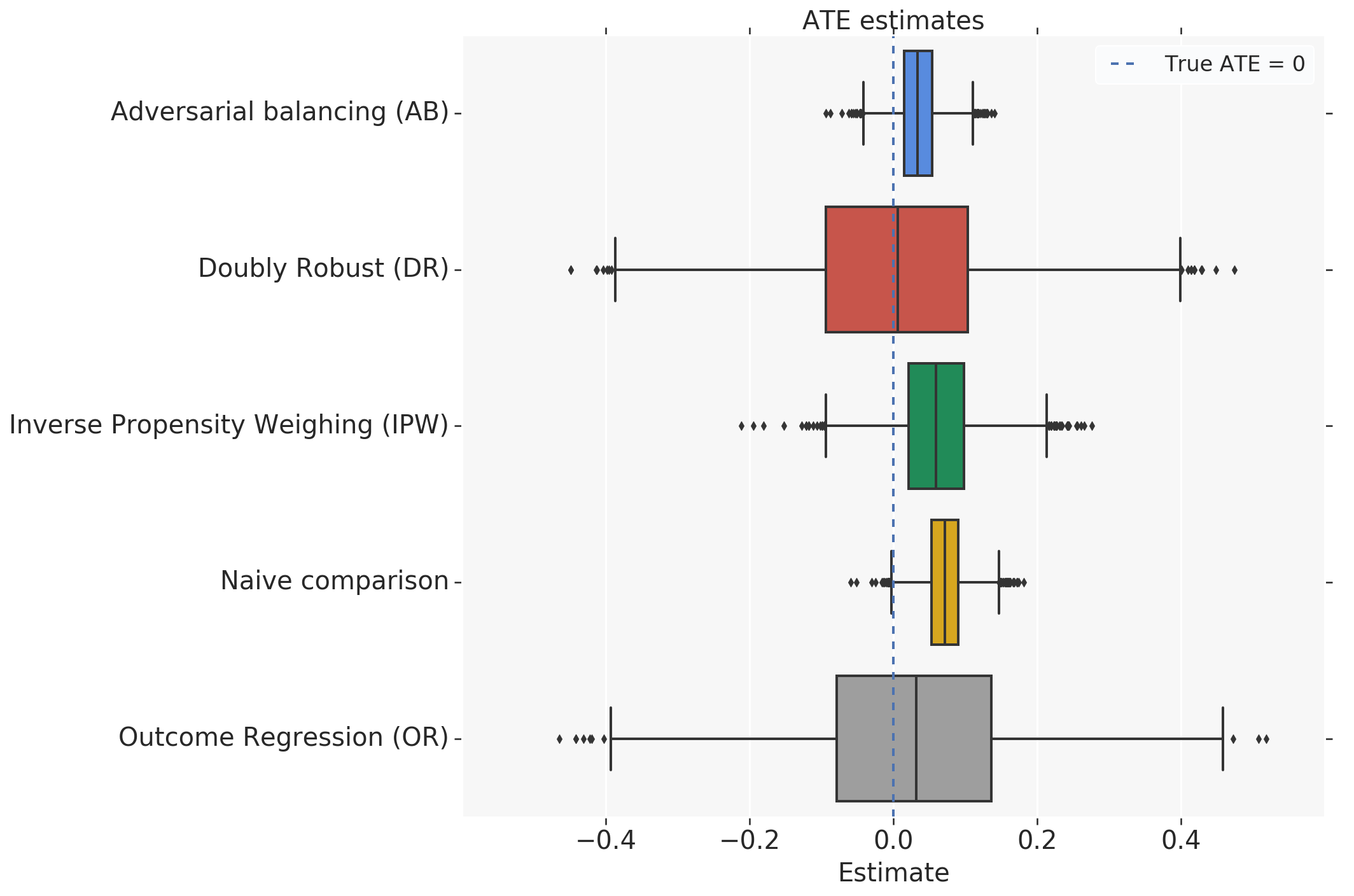}
\caption{Performance of different ATE estimators when noisy confounders are observed. Increase in variances of OR and DR, AB retains performance characteristics.}
\label{figure_sim_ATE_2}
\end{figure}

In scenario 2, we relax our assumption on the ability to observe the true confounders. Instead, we assume that we can only observe transformations of the true confounders as given in Eq.~\eqref{eq.z1}-\eqref{eq.z2}. This corresponds  to a case where the researcher cannot directly observe $X$, but can still access a set of variables that are related to the confounders themselves (and thus will likely misspecify the model). Results and boxplots are given in Table~\ref{table_ate_results_2} and Figure~\ref{figure_sim_ATE_2}, respectively. 

\begin{table*}\centering
\ra{1.3}
\begin{tabular}{llll}
\toprule
\cmidrule(r){1-2}
 Estimator & Bias & MAE & MSE \\
\midrule
 AB & \num{3.4e-2} & \num[math-rm=\mathbf]{3.4e-2} & \num[math-rm=\mathbf]{1.8e-3}\\
 DR & \num[math-rm=\mathbf]{6.2e-3} & \num{9.6e-2} & \num{2.0e-2}\\
 IPW & \num{5.9e-2} & \num{6.0e-2} & \num{6.6e-3}\\
 Naive comparison & \num{7.1e-2} &  \num{7.1e-2} & \num{5.8e-3} \\
 OR & \num{3.0e-2} & \num{1.0e-1} & \num{2.4e-2} \\
\bottomrule
\end{tabular}
\caption{Performance of different ATE estimators when noisy confounders are observed. Estimators are Adversarial Balancing (AB), Doubly Robust (DR), Inverse Propensity Weighing (IPW), Naive mean comparison and Outcome Regression (OR). In contrast to Table~\ref{table_ate_results}, the performance of OR drops significantly while AB continues to provide a balance between variance and bias.}
\label{table_ate_results_2}
\end{table*}

Our first observation is the dramatic drop in the performance of OR. The estimator performs the worst in both bias and MSE.  
DR achieves the best performance in this case with respect to bias, however it once again suffers from large variance. The AB estimator seems to provide a balance between bias and variance.

\subsection{Sentiment Average Treatment Effect on Treated and Responded (ATETR)}
Similar to the ATE case, we have implemented five different estimators under two different scenarios. 

In scenario 1, we assume that we can observe the actual values of the confounding factors, \emph{i.e.}, $X_1,X_2$. Our first estimator is a naive mean comparison among respondents. The naive estimator will return the same estimate for ATE and ATETR. We implemented outcome regression (OR)~\cite{doi:10.1111/j.1541-0420.2005.00377.x}, inverse propensity weighing (IPW)~\cite{kang2007}, covariate control (CC), an adversarial balancing based estimator\footnote{We use logistic regression for the discriminator.} (AB)~\cite{2018arXiv181007406O}, and an entropy balancing\footnote{We utilized first moment based balancing for EB estimator.} based method (EB)~\cite{hainmueller_2012}.
The main difference between ATE and ATETR estimation is the latter utilizes data among respondents only. 

\begin{table*}\centering
\ra{1.3}
\begin{tabular}{llll}
\toprule
\cmidrule(r){1-2}
 Estimator & Bias & MAE & MSE \\
\midrule
 AB & \num{3.2e-2} & \num{3.3e-2} & \num{1.7e-3}\\
 CC & \num{7.7e-3} & \num{1.9e-2} & \num{7.8e-4} \\
 EB & \num[math-rm=\mathbf]{4.3e-3} & \num[math-rm=\mathbf]{1.8e-2} & \num[math-rm=\mathbf]{7.1e-4}\\
 IPW & \num{7.7e-3} & \num{1.9e-2} & \num{1.0e-3}\\
 Naive comparison & \num{7.1e-2} &  \num{7.1e-2} & \num{5.8e-3} \\
 OR & \num{7.8e-2} & \num{7.8e-2} & \num{6.8e-3} \\
\bottomrule
\end{tabular}
\caption{Performance of different ATETR estimators when true confounders are fully observed. Estimators are Adversarial Balancing (AB), Covariate Control (CC), Entrophy Balancing (EB), Inverse Propensity Weighing (IPW), Naive mean comparison and Outcome Regression (OR). EB has the best performance followed by simple CC estimator.}
\label{table_atetr_results}
\end{table*}

\begin{figure}[]
  \centering
\includegraphics[width=1\textwidth]{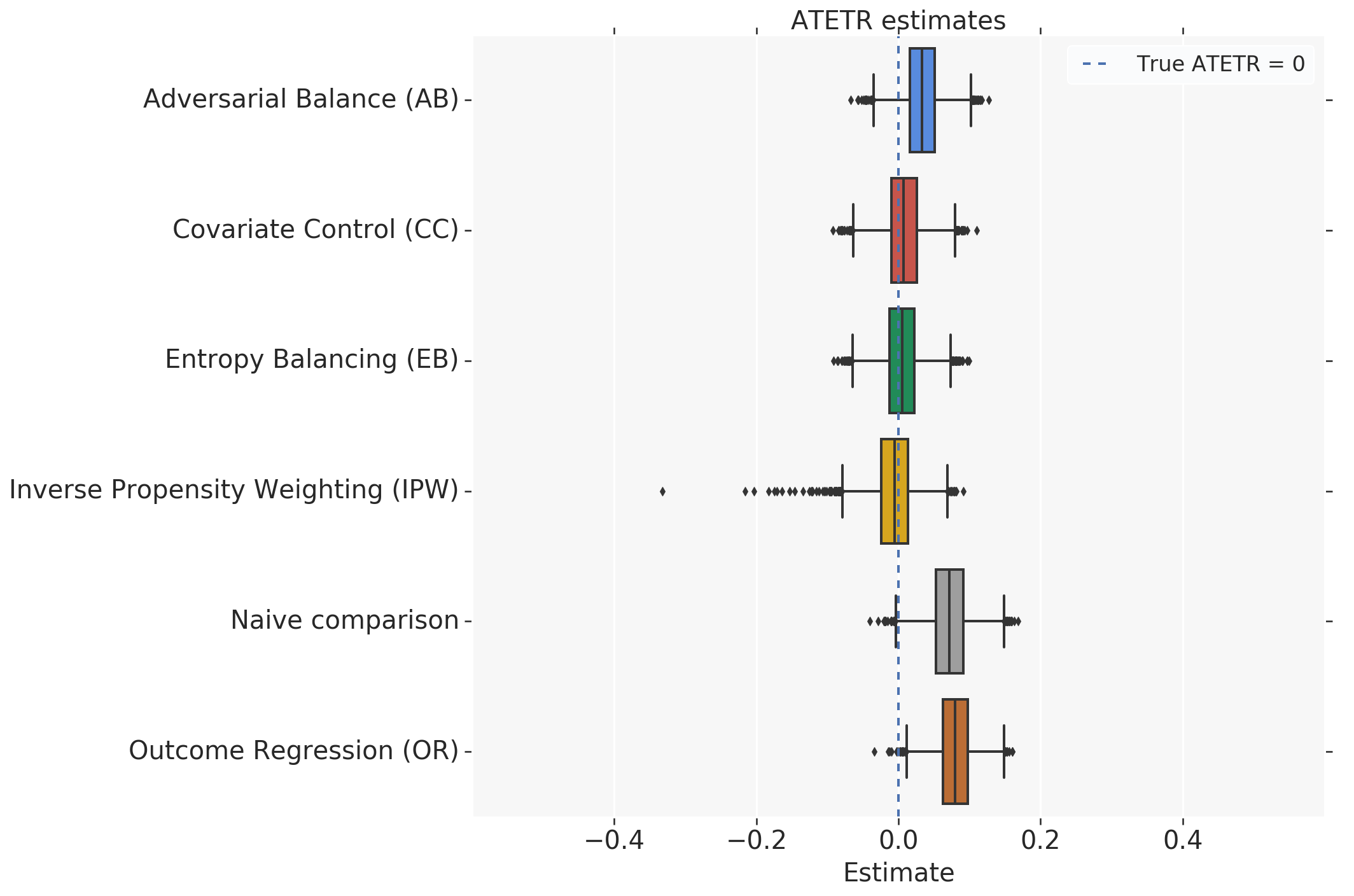}
\caption{Performance of different ATETR estimators when true confounders are fully observed. CC and EB outperforms AB in contrast to ATE.}
\label{figure_sim_ATETR_1}
\end{figure}

Results and boxplots are given in Table~\ref{table_atetr_results} and Figure~\ref{figure_sim_ATETR_1}, respectively. In our tests, the OR estimator has the largest bias. The estimator is trained using control-assigned survey respondents, and counterfactuals for the treatment assigned respondents are generated using the trained model. In our setting, the OR estimator consistently underestimates counterfactuals for the treatment-assigned respondents, resulting in a bias that is comparable with the naive estimator. CC, EB and IPW have small biases; EB and CC estimators have the best MSE performance across all estimators.

\begin{table*}\centering
\ra{1.3}
\begin{tabular}{llll}
\toprule
\cmidrule(r){1-2}
 Estimator & Bias & MAE & MSE \\
\midrule
 AB & \num{4.1e-2} & \num{4.1e-2} & \num{2.5e-3} \\
 CC & \num{2.6e-2} & \num{2.8e-2} & \num{1.5e-3} \\
 EB & \num[math-rm=\mathbf]{1.7e-2} & \num[math-rm=\mathbf]{2.3e-2} & \num[math-rm=\mathbf]{1.2e-3} \\
 IPW & \num{-3.3e-1} & \num{3.1e-2} & \num{5.8e-1}\\
 Naive comparison & \num{7.1e-2} & \num{7.1e-2} & \num{5.8e-3}\\
 OR & \num{1.2e-1} & \num{1.2e-1} & \num{1.6e-2} \\
\bottomrule
\end{tabular}
\caption{Performance of different ATETR estimators when noisy confounders are observed.  Estimators are Adversarial Balancing (AB), Covariate Control (CC), Entrophy Balancing (EB), Inverse Propensity Weighing (IPW), Naive mean comparison and Outcome Regression (OR). Similar to the results in Table~\ref{table_atetr_results}, EB outperforms across all measures. Similar to Table~\ref{table_atetr_results}, simple CC estimator provides a comparable performance to EB.}
\label{table_atetr_results_2}
\end{table*}

\begin{figure}[]
  \centering
\includegraphics[width=1\textwidth]{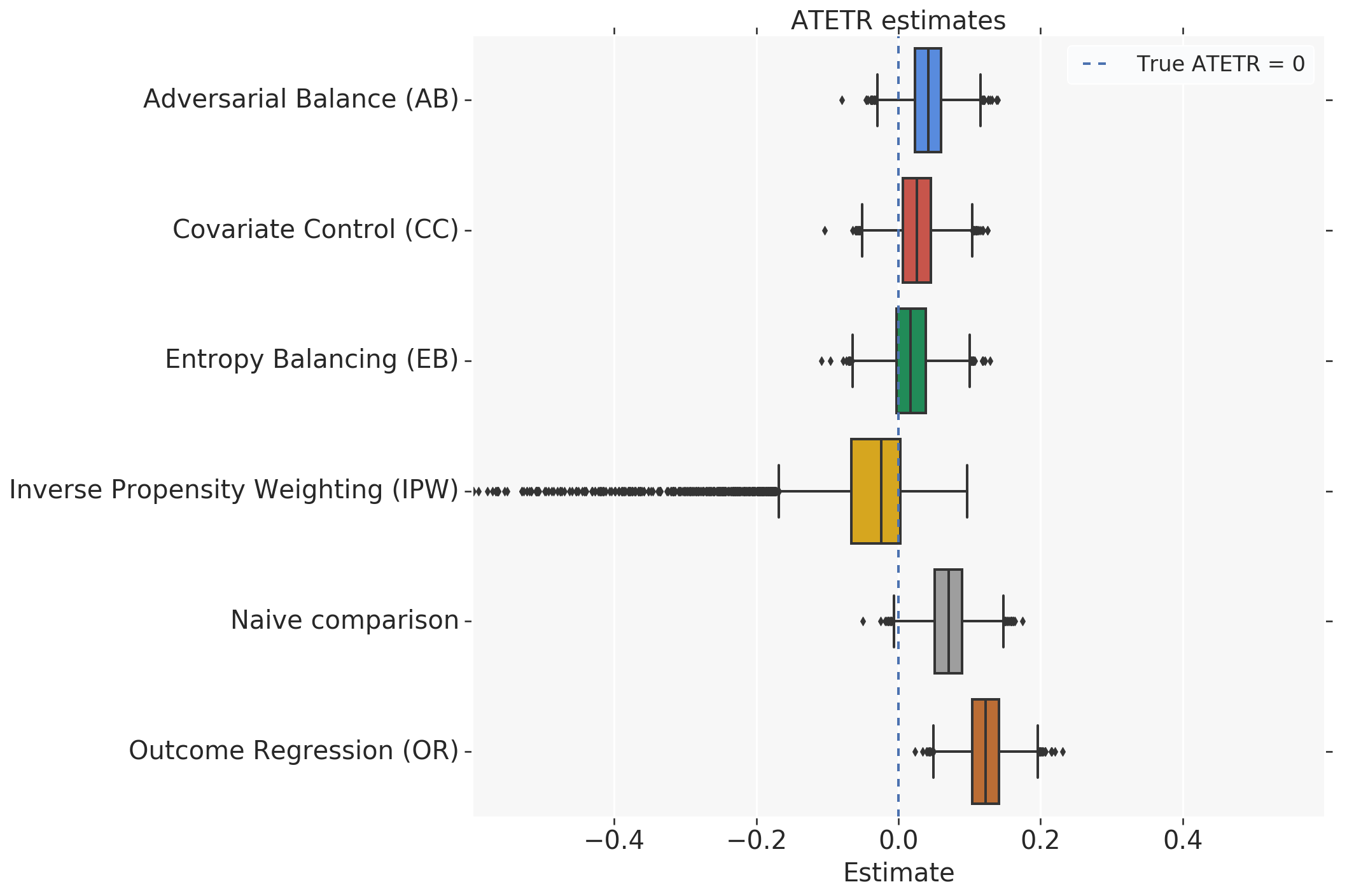}
\caption{Performance of different ATETR estimators when noisy confounders are observed.  Estimators are Adversarial Balancing (AB), Covariate Control (CC), Entrophy Balancing (EB), Inverse Propensity Weighing (IPW), Naive mean comparison and Outcome Regression (OR). Similar to the results in Table~\ref{table_atetr_results}, EB and CC outperform across all measures. IPW suffers from large variance, OR performs worse than naive estimator.}
\label{figure_sim_ATETR_2}
\end{figure}

In scenario 2, we relax our assumption on our ability to observe the true confounders, and we assume that we can only observe transformations of the true confounders as given in Eq.~\eqref{eq.z1}-\eqref{eq.z2}. Results and boxplots are given in Table~\ref{table_atetr_results_2} and Figure~\ref{figure_sim_ATETR_2}, respectively. 

CC suffers from a dramatic drop in performance when compared to the fully observed covariates scenario. We also observe dramatic bias and higher levels of variance in IPW as well as a large number of estimates with extreme values. The performance of the AB estimator seems comparable between the two scenarios. While the EB estimator suffers from increased bias, its bias and MSE are nevertheless the lowest.

\section{Conclusion}
In this paper, we focused on a scenario where user reported sentiment is used as a success metric for a fully randomized A/B test. Sentiment is assumed to be collected via a survey, and users have the option to not respond to the survey. 

In the case of Average Treatment Effect (ATE), we showed that our problem can be broken down into two independent missing data problems. In other words, we can focus on treatment and control sets separately, and estimate population-level sentiment for both groups. Then we show that a simple difference between the two quantities provides a consistent estimator of the ATE. This also enables parallel execution for more efficient computation. 

We simulate the problem with both fully observed and transformed confounders, observing significant performance decreases between two settings, highlighting the potential sensitivity of causal inference techniques to model misspecification. The adversarial balancing estimator seems to provide a desirable balance between bias and variance in the transformed confounders scenario. 

In the case of Average Treatment Effect on Treated and Responded (ATETR), we showed that one can focus on respondent data only. Our problem can be mapped onto a standard causal analysis setting, \emph{i.e.}, matching respondents in the control group to respondents in the treatment group using the observed covariates. This enabled us to leverage existing techniques in the causal analysis literature. 

Similarly to ATE, we observe significant performance decreases between the two ATETR settings. In particular, IPW suffers from dramatic performance degradation (potentially due to extreme values of predicted propensities). Interestingly, a simple regression-based covariate control (CC) was one of the best estimators for the fully observed case, suggesting more complicated techniques do not necessarily provide superior performance.

Our novel framework combined with simple regression or balancing techniques has the potential to be integrated with existing experimentation platforms to provide robust, efficient, and automatic impact calculations for sentiment metrics.

\bibliography{bib_1}
\bibliographystyle{acm}

\appendix
\section{Proof of Lemma~\ref{lemma_ATETR}}\label{proof_lemma_2}
First recall that ATETR is defined as $E[Y(1)-Y(0)| T=1, \Delta(T) = 1]$ in Equation~\eqref{main2_eq}. By the linearity of the expectation operator:
\begin{align*}
ATETR  = E[Y(1)| T=1, \Delta(T) = 1] - E[Y(0)| T=1,\Delta(1) = 1].
\end{align*}
A simple average over the set of users who are assigned to the treatment ($T=1$) and responded to the survey $\Delta(1) = 1$ is a consistent estimator of the first quantity, \emph{i.e.}, $E[Y(1)| T=1, \Delta(T) = 1]$. This follows from the fact that assignment is random. 

If we focus on the second term, the following set of equalities hold:
\begin{align}
E[Y(0)| T=1, \Delta(T) = 1] &= E[Y(0)|\Delta(1) = 1]\label{ap1} \\
&= E[E[Y(0)| \Delta(1) = 1,  \Delta(0)=1] | \Delta(1) = 1 ]\label{ap2}\\
&= E[E[E[Y(0)|\Delta(1) = 1, \Delta(0)=1,  X] | \Delta(1) = 1, X ]] \label{ap4}\\
&= E[E[E[Y(0)| \Delta(0)=1, X] | \Delta(1) = 1, X ]]. \label{ap5}
\end{align}
Equation~\eqref{ap1} follows from the fact that treatment/control groups are randomly assigned, Equation~\eqref{ap2} and ~\eqref{ap4} hold due to the law of iterated expectations, Equation~\eqref{ap5} follows from Assumption~\ref{assumption_2}.

\section{Data Generating Process for simulations in Section~\ref{simulations}}\label{appendix_DGP}
Given our causal graph in Figure~\ref{figure_simulated_causal}, we draw Monte Carlo simulations using the following data generating process (DGP):
\begin{align*} 
X_1 &\sim  \text{Exp}(\lambda=2), ~ X_2 \sim  \text{Exp}(\lambda=3)  \\
S &= 2X_1 - 1.5X_2 \\
Y &\sim \text{Bin}\left(\frac{1}{1 - e^{-S}}\right) \\
R &= X_1 T - X_2 T - X_2 (1 - T) - 2 \\
\Delta(T)  &\sim \text{Bin}\left(\frac{1}{1 - e^{-R}}\right)
\end{align*}
where $\text{Exp}(\lambda)$ is an exponential distribution with rate $\lambda$, $\text{Bin}(\theta)$ is a binomial distribution with probability $\theta$, $T$ is a treatment indicator ($T=1$ for treatment and $T=0$ for control), $Y$ is the sentiment, and $\Delta(T)$ is an indicator of whether a response was received under treatment $T$. Our DGP satisfies both Assumptions~\ref{assumption_1} and~\ref{assumption_2} since $X=[X_1,X_2]$ confound both response rates and sentiment. True sentiment $Y$ is the same across treatment and control since parameter $S$ is independent of the treatment indicator $T$. Response rates, however, are a function of both the confounders $X$ and the treatment indicator $T$. Therefore, the observed sentiment for the two groups is different. 

In some of our simulations, we relax the observability on the true confounders $X_1,X_2$ to study the effects of model misspecification. We instead assume that we can observe functions of the confonfounders, $Z_1, Z_2$ where:
\begin{align} 
Z_1 &= e^{X_1/2} \label{eq.z1}\\ 
Z_2 &= X_2^2 - X_2 + X_1 * X_2\label{eq.z2}
\end{align}
This corresponds to a scenario where a researcher cannot directly observe the confounders but has access to a set of proxy variables for the confounders.

\end{document}